\documentclass[conference]{IEEEtran}
\IEEEoverridecommandlockouts

\usepackage{cite}
\usepackage{amsmath,amssymb,amsfonts}
\usepackage{algorithmic}

\usepackage{caption}
\usepackage{listings}
\usepackage{float}

\newfloat{listing}{htp}{lst}
\floatname{listing}{Listing}

\usepackage{graphicx}
\usepackage{textcomp}
\usepackage{xcolor}
\usepackage{url}
\def\BibTeX{{\rm B\kern-.05em{\sc i\kern-.025em b}\kern-.08em
    T\kern-.1667em\lower.7ex\hbox{E}\kern-.125emX}}
    
\begin{document}

\title{Quantum Resource Management in the NISQ Era: Challenges, Vision, and a Runtime Framework}

\author{

\IEEEauthorblockN{1\textsuperscript{st} Marcos Guillermo Lammers}
\IEEEauthorblockA{\textit{LIFIA, Facultad de Informática/CICPBA} \\
\textit{Universidad Nacional de La Plata}\\
La Plata, Argentina \\
\url{https://orcid.org/0009-0007-3954-2252}}

\and

\IEEEauthorblockN{2\textsuperscript{nd} Federico Hernán Holik}
\IEEEauthorblockA{\textit{Instituto de Física La Plata}\\
\textit{CONICET/UNLP}\\
La Plata, Argentina \\
\url{https://orcid.org/0000-0002-6776-5281}}

\and

\IEEEauthorblockN{3\textsuperscript{rd} Alejandro Fernández}
\IEEEauthorblockA{\textit{LIFIA, Facultad de Informática/CICPBA} \\
\textit{Universidad Nacional de La Plata}\\
La Plata, Argentina \\
\url{https://orcid.org/0000-0002-7968-6871}}

}

\maketitle

\begin{abstract}
Quantum computers represent a radical technological advancement in the way information is processed by using the principles of quantum mechanics to solve very complex problems that exceed the capabilities of classical systems. However, in the current NISQ era (Noisy Intermediate-Scale Quantum devices), the available hardware presents several limitations, such as a limited number of qubits, high error rates, and reduced coherence times. Efficient management of quantum resources,  both physical (qubits, error rates, connectivity) and logical (quantum gates, algorithms, error correction), becomes particularly relevant in the design and deployment of quantum algorithms. In this work, we analyze the role of resources in the various uses of NISQ devices today, identifying their relevance and implications for software engineering focused on the use of quantum computers. We propose a vision for runtime-aware quantum software development, identifying key challenges to its realization, such as limited introspection capabilities and temporal constraints in current platforms. As a proof of concept, we introduce Qonscious, a prototype framework that enables conditional execution of quantum programs based on dynamic resource evaluation. With this contribution, we aim to strengthen the field of Quantum Resource Estimation (QRE) and move towards the development of scalable, reliable, and resource-aware quantum software.
\end{abstract}

\begin{IEEEkeywords}
Quantum Computing, Quantum Resource Management, Quantum Software Engineering, Quantum Resource Estimation (QRE)
\end{IEEEkeywords}

\section{Introduction}

Quantum computing is an interdisciplinary field of research and development that promises relevant future applications in areas such as cybersecurity, solving optimization problems in industry and finance, and simulating molecules for the chemical and pharmaceutical industries.

Quantum computers already exist, some of which are commercially available. In this work, we will distinguish between the concept of a ``fault-tolerant quantum computer'' (FTQC) and the currently available devices, which we will refer to as noisy or imperfect. Currently, no FTQC is robust enough to perform commercially relevant applications, such as applying Shor’s algorithm \cite{shor1994} to break a cryptographic key. Experts refer to today's devices with the acronym ``NISQ'' (Noisy Intermediate-Scale Quantum). Although the advances in recent years have been significant \cite{sycamore2019, Zuchongzhi2024}, most experts agree that we are several years away from the era of FTQC and relevant commercial applications. Some authors estimate that we could be decades away from the era of FTQC capable of solving commercially relevant problems.

The NISQ era is a term coined by John Preskill \cite{Preskill2018}. In this context, the author refers to computers having between 50 and 100 qubits. He describes them as computers limited by noise to unreliable and reduced quantum circuits. However, they still may be helpful in experiments and research related to quantum physics. Nowadays, there are publicly accessible, gate based quantum computers with 1000+ qubits. Still, noise remains a complex problem to address. Some currently available computers based on ion traps have less noisy qubits and higher fidelity gates, but they have around 50 physical qubits. In addition, the dispersion of existing hardware, the lack of standards, the high cost of development, and the technical difficulties in scalability make it unfeasible to apply algorithms and quantum circuits of commercial interest. There are many promises for fault-tolerant computers in the future \cite{iqm-roadmap, ibm-roadmap, quandela-roadmap, microsoft2025}. Many states and companies are working on their development. So far, despite advancements, fault-tolerant computers with a reasonably large number of qubits have not become a reality. However, the currently available NISQ devices are increasingly used for research and development purposes. A software engineering approach that considers their limitations will allow us to facilitate their use during the NISQ era and better prepare them for potential technological upgrades in the coming years.

Although there is no widespread commercial use of NISQ devices, it is essential to highlight that a community involving various types of users has progressively developed in recent years. These are concentrated in research teams from public institutions and companies. The research conducted is diverse and can be classified (at least) into three groups. The first group focuses on improving existing devices by detecting and correcting their errors and identifying the steps to follow in their future development. The second group refers to the early positioning of companies, states, and academic groups regarding the resolution of future problems (an example is the development of research teams in pharmaceutical companies seeking medium-term applications). Finally, a third group focuses on studying whether it is possible to gain some advantage, either commercial or pure research, from using NISQ devices. Naturally, the distinction between these three groups is blurred, and there is significant overlap between the different tasks. These projects are evidence of a considerable community of NISQ device users for whom the development of software tools to achieve their research and development goals has become a daily challenge. This implies the need to invest considerable resources in developing such tools. Given the advances of recent years, it is essential to remember that this situation could extend for several years and even decades.

In this context, we can precisely formulate the question that we will address in this work: What are the implications for quantum software engineering of the fact that users of quantum computers must deal with NISQ devices, which have limited resources and are subject to errors? The relevance of this question is linked to another fact: most of the research in the field known as Quantum Software Engineering focuses on FTQC. That is, it concentrates on the study of devices to come without considering the challenges posed by the research carried out by users of NISQ devices. This work aims to identify and systematize some of the development practices relevant for software engineering aimed at using NISQ devices.

This article is organized as follows: Section \ref{s:Ingeniería_de_software} discusses the role of resources in classical software engineering, highlighting parallels with quantum computing. Section \ref{secction:rcq} introduces the concept of quantum resources in NISQ devices, distinguishing between physical and logical layers. Section \ref{section:related-work} surveys existing efforts in quantum resource estimation and benchmarking, identifying their current focus on fault-tolerant architectures. Section \ref{s:Recursos_en_NISQ} presents a vision for runtime quantum resource management in the NISQ era and motivates the need for introspection and adaptivity. It also introduces Qonscious, a prototype framework that enables conditional quantum execution based on dynamic resource checks. Section \ref{section:practical-resource-management} analyzes key implementation challenges for such a runtime layer, including API limitations, extensibility, and temporal constraints. Finally, Section \ref{section:conclusions} offers conclusions, outlines future work, and advocates for greater developer awareness around resource-conscious quantum software development.

% Secciones 
\section{The role of resources in software engineering}
\label{s:Ingeniería_de_software}

Software engineering, as a discipline, has among its main objectives the efficient management of available resources, recognizing that these are finite and that the success of projects depends on them. It is a discipline that deals with the different phases of the software production process, from the early stages of the system to be developed to its maintenance once it is in use \cite{sommerville2016}.

To execute any software, it is necessary to consider different types of resources, such as processing time, memory, files, and input/output devices. The software can require these resources at the moment of its creation or during its execution \cite{silberschatz2006}. This means that, from the beginning, engineers must organize, among other things, the use of hardware and software resources to prevent bottlenecks and ensure optimal performance. As Sommerville mentions, in traditional software engineering, the challenge is to create techniques for generating reliable software that has enough adaptability to manage heterogeneity, which includes the ability to adapt to the limitations of existing hardware.

Hardware resources directly influence decisions such as selecting software architectures and performance optimization on any device. This is no different in the use of quantum computers. The hardware must be used efficiently to meet performance and reliability requirements. The main challenge is to find the balance between scarce resources and project objectives, ensuring that the software is efficient, scalable, and capable of operating within the available physical limitations. To verify this, we need to reliably understand the available hardware and software requirements.

\section{Quantum Computer Resources}
\label{secction:rcq}

We classify the resources of a quantum computer according to their physical and logical nature. Physical resources reflect constraints and capabilities determined by the hardware design—such as the number of qubits, error rates, coherence time, entanglement capacity, connectivity, and gate fidelity—which are often vendor-specific and may be partially abstracted from the developer. Logical resources are software-visible abstractions that describe what can be expressed and executed on the hardware—such as supported gate sets, circuit depth, error correction mechanisms, and available measurements.

The potential advantage of quantum computers depends on the type and quality of their physical resources. These enable the quantum effects that distinguish quantum from classical computation. Logical resources, exposed through software platforms, allow users to access and coordinate those capabilities. Physical resources define what is possible to compute, while logical resources influence how much effort developers require to realize that potential. While the expectation is that physical constraints will become less visible to developers as we move toward fault-tolerant quantum computing (FTQC), in the current NISQ era they remain a critical and immediate concern. In this context, software engineering must address both layers: the variability and limitations of the physical substrate, and the abstractions available to manage them.

In quantum computing, a large number of physical qubits is often considered an advantage as it means having greater data processing capacity. In theory, a larger number of physical qubits would allow for more logical qubits, on which quantum error correction protocols could be implemented. It is important to emphasize that, although many physical qubits are theoretically desirable, their quality must also be considered. Physical qubits with a high error rate will not be functional, even if they are numerous. This fact underscores the importance of the quality of qubits as a key resource \cite{Preskill2018}. The error rate is closely tied to the quantum computer (QC) type. A low error rate must be achieved to execute algorithms, obtain reliable results, and ensure the system is scalable.

The coherence time tells us how long a qubit can maintain its quantum state before decoherence degrades it and the information is lost. For a quantum algorithm to execute correctly, the coherence time must be long enough to allow all quantum circuit operations to be carried out. If it is too short, the qubits may lose their superposition and entanglement properties, and the algorithm's results will be incorrect and unreliable. Many algorithms, such as Shor's \cite{shor1994} or Grover's \cite{grover1996}, depend on these properties to achieve their advantage over classical methods. A weak level of entanglement means that the algorithm will not have any quantum advantage or produce incorrect results.

Native gates are basic operations of quantum hardware. These are implemented directly in the physical architecture of the quantum processor and are specific to the type of technology used (for example, superconducting qubits or trapped ions). They are the basic building blocks for constructing more complex quantum algorithms \cite{krantz2019}. Clifford gates (such as Hadamard, Pauli-X, Pauli-Z, and CNOT) are basic quantum operations used to manipulate and entangle qubits. They can be efficiently simulated on classical computers \cite{nielsen2010}, which implies that a device that only uses Clifford gates and measures and prepares states in the computational basis does not exploit any quantum advantage \cite{Gottesman2004}. If the T gate is added to the Clifford group, the Clifford+T set is obtained, which is universal. A device whose native gates form a universal set can perform, in principle, any quantum computation. Imperfections in NISQ devices impose severe restrictions on the quantum advantage. This implies that in software engineering, the implementation of these gates must be carefully considered to achieve resource optimization (\cite{bravyi2005}).

The connectivity of qubits is their ability, determined by the hardware topology, to establish entangling interactions with each other. We say that two qubits are connected if it is possible to apply an entangling native gate between them. In systems with limited connectivity, qubits can only interact with their nearest neighbors, which requires extra operations (such as state swaps with Swap gates) to entangle non-adjacent qubits. This increases the complexity of the circuit and the execution time of the algorithms, and therefore the overall error rate. Some architectures have full connectivity. This means it is possible to apply an entangling gate directly between any pair of qubits chosen, as with ion trap-based processors (IonQ, AQT). On the other hand, devices based on superconducting qubits (IBM, Google, IQM) tend to have limited connectivity.

The noise level consists of random and uncontrolled fluctuations in the physical parameters that interact with the qubits. These fluctuations can originate from various sources, such as thermal noise (voltage and current variations), amplitude or phase fluctuations in the oscillators that generate control pulses, or fluctuating electric and magnetic fields in the local environment of the qubit, such as in metallic surfaces or substrate-metal interfaces. These disturbances cause unwanted changes in the qubit's parameters, leading to decoherence and reducing the fidelity of quantum operations. Noise can be classified into two main types: systematic noise, which consists of predictable and repeatable errors, such as an incorrect rotation of a microwave pulse due to imperfect calibration, and stochastic noise, which consists of random and unpredictable fluctuations, such as those caused by thermal noise or electromagnetic interference \cite{krantz2019}.
 
\section{Quantum Resource Management Today}
\label{section:related-work}

The current approaches to studying and managing resources in quantum computers can be classified into those focusing on resource estimation and those focusing on benchmarking. The former aims to determine the resource requirements of a specific quantum algorithm or system. The latter seeks to establish objective evaluation standards to compare quantum technologies.

\subsection{Quantum Resource Estimators}
\label{qre}
Several projects are currently being developed within the research area known as Quantum Resource Estimators (QRE). These projects focus on estimating and optimizing the resources required primarily for fault-tolerant universal quantum computers.

Microsoft's resource estimator (Azure Quantum Resource Estimator) calculates the execution time and total number of qubits based on an estimated cost of the number of gates required for quantum circuits \cite{AzureQRE2023}.

Google developed Qualtran, which is currently in beta. Qualtran simulates and tests algorithms using abstractions and data structures to generate diagrams with information and tabulate resource requirements automatically. They offer a standard library of building blocks for cost-minimizing builds (\cite{qualtran2024}).

Zapata AI developed BenchQ as part of the DARPA Quantum Benchmarking Program \cite{darpa2023}. BenchQ provides tools to estimate the hardware resources needed for fault-tolerant quantum computing. It includes a graphical state compiler, distillation factory models, decoder performance evaluations, an ion trap architecture design, and implementations of specific quantum algorithms, among other features. Underneath, it uses the tools offered by Microsoft Azure Quantum Resource Estimation (QRE) \cite{benchq2025}.

M. Suchara and collaborators \cite{QuRE2013} developed the QuRE tool. They describe it as an estimation application that calculates the cost of practical implementations of quantum circuits across various physical quantum technologies, specifically for fault-tolerant computer coding.

The Munich Quantum Toolkit (MQT) offers a set of benchmarks to evaluate software tools and design automation in quantum computing called MQT Bench\label{MQT Bench}. They use pre-designed quantum circuits at four levels of abstraction, from the algorithmic level to the hardware-dependent level. They have prior analyses on the suitability of execution across different computers from IBM, Rigetti, IonQ, OQC (Oxford Quantum Circuits), and Quantinuum (considering their native gates). They provide support for access via a web interface and a Python package. They use the Qiskit and/or TKET compiler and their default optimizers. They have five sets of native gates and seven quantum devices with capabilities ranging from 8 to 127 qubits. They aim to improve comparability, reproducibility, and transparency in evaluating quantum software tools \cite{MQTBenchBenchmarking2023}.

The work done by Saadatmand \cite{FTQCs2024} focuses on estimating the resources required for FTQC. The authors use a methodology based on graph-state compilation to assess resource requirements in FTQC systems. They focus on superconducting architecture and analyze how connectivity between modules, latency, and other key factors impact the total resources needed to implement large-scale quantum algorithms. They provide a tool and a methodological framework to evaluate and optimize resources in quantum architectures.

In a work conducted for DARPA and NASA, Mozgunov, Marshall, and Anand \cite{DarpaNasa2024} describe two applications in which their preferred approach is the quantum simulation of open systems. The authors propose algorithmic optimizations by selecting specific parameters for resource estimation in problems where the execution time is considerable. These optimizations leverage translational invariance, which is the property of a system or mathematical function that remains unchanged when shifted or translated in space, simplifying calculations and taking advantage of periodically repeating structures, as well as parallelism in the application of the T-gate, thus reducing the required resources. These optimizations are designed to be implemented on FTQC. Resource estimates for these systems are centrally dependent on the T-state footprint, meaning the resources required to generate and maintain T-states.

\subsection{Benchmarking}
\label{benchmarking}

In the projects described in the previous section, the primary approach for resource estimation depends on the algorithm, its functionality, and purpose, or the prior data we have about the quantum computer provided by the service provider. There are also many benchmarking projects to evaluate the hardware based on different factors \cite{BenchmarkingQuantumComputers2025}. This also influences the results of tests, announcements, and conclusions, both by the researchers and their competitors.

In 2018, the Google team designed a benchmarking method for quantum supremacy \cite{boixo2018,bravyi2005}. This method is based on theoretical studies about the difficulty of sampling from random quantum circuits, the computational hardness of their classical simulation, and its implications for demonstrating quantum supremacy, establishing connections with computational complexity and separations between quantum and classical models \cite{aaronson2017}.

The experimental implementation of sampling on random quantum circuits led Google to announce that the Sycamore processor took approximately 200 seconds to sample a million times an instance of a quantum circuit. They indicated that the same task for a state-of-the-art classical supercomputer would take about 10,000 years and that this quantum supremacy also marked the era of noisy intermediate-scale quantum (NISQ) technologies. Additionally, they claimed that this benchmarking ensured an immediate application in generating certifiable random numbers \cite{sycamore2019}. Shortly afterward, IBM \cite{pednault2019} proposed a much more efficient classical hardware simulation to perform the same execution in 2.5 days, not in 10,000 years.

This experience demonstrates that one must carefully select the tests and benchmarks to draw correct conclusions.

Not every benchmarking is helpful in every case. In \cite{BenchmarkingQuantumComputers2025}, multiple available methods are described, detailing which are more suitable for the NISQ era, others for the fault-tolerant stage, low level, high level, and how this will necessarily change as technological progress advances. They also describe six fundamental properties for achieving a good benchmark:

\begin{enumerate}
    \item They must be well-founded, justifying the performance metrics.  
    \item The procedure must be unambiguous. Any step not specified in the procedure should be an intentionally configurable parameter.
    \item It should not be possible to manipulate the configurable parameters in its implementation to obtain misleading results.
    \item The results should not be corrupted when the evaluated quantum computer experiences small or unforeseen errors.
    \item They must be efficient and not consume many resources.
    \item Finally, the benchmark should be technology-independent; it should only be specific if what is being measured is particular to that architecture.
\end{enumerate}

While most existing work on quantum resource estimation and bechmarking focuses on tooling for hardware comparison or algorithm cost modeling, very few efforts address the role of resource awareness in developer workflows and education. We believe this is a crucial missing link: developers need tools not only to measure resources but to understand and internalize their impact. Our proposal aims to bridge this gap by embedding resource reasoning directly into the software development cycle, encouraging developers to reflect on quantum resource requirements as a central part of their design process.

\section{A Vision of Quantum Resource Management in the NISQ Era}
\label{s:Recursos_en_NISQ}

From the perspective of Software Engineering, determining the available resources in current quantum computers, regardless of the platform, and reliably managing them correctly remains an unresolved challenge.

The testing techniques mentioned in \ref{benchmarking} and the libraries or tools detailed in \ref{qre} can be helpful in different cases. However, aside from mostly only functioning with FTQC, their application must be done before execution, verifying requirements. This represents a static view of the use of quantum computers that does not consider the particularities of how they work or the technological development stage of utility we are currently in. In this last point, we can make an exception with MQT Bench \ref{MQT Bench}, which is based on previously obtained data, but not updated when executing the algorithm. The connectivity of the qubits may remain the same if the hardware topology does not change. Still, their level of entanglement, coherence time, or fidelity could change from one moment to another.
    
\begin{figure}
    \centering
    \includegraphics[width=0.9\linewidth]{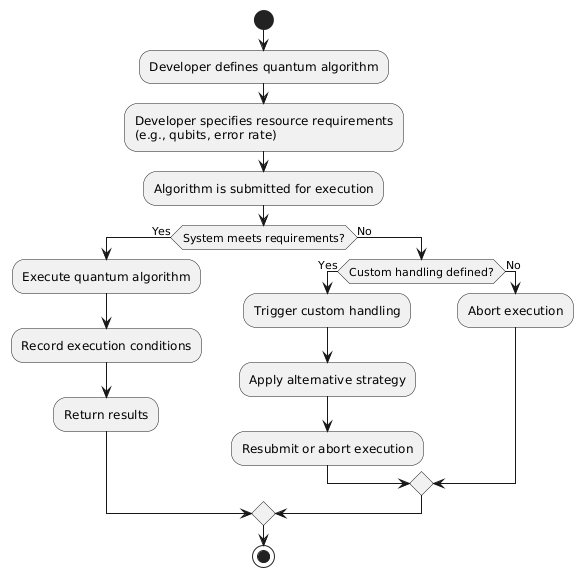}
    \caption{Intended development workflow supported by the proposed layer}
    \label{fig:workflow}
\end{figure}

We claim that the current state of quantum hardware and its typical usage scenarios demand enhanced capabilities from programming platforms. In particular, developers should be able to programmatically express resource constraints and monitor resource availability and usage at runtime. We propose that quantum software development platforms incorporate a dynamic runtime layer for resource management and introspection. This layer should operate at execution time to assess whether the necessary quantum resources are currently available and adequate for the intended computation. Quantum software development would take the form depicted in the workflow presented in Fig. \ref{fig:workflow}. Unlike existing tools, which typically rely on static or pre-execution analysis, the proposed layer must address the dynamic behavior of NISQ devices, where critical parameters such as fidelity, decoherence, and qubit connectivity may fluctuate between executions, even on the same hardware. 

Countable and easily identifiable resources in a pre-execution instance include the number of qubits needed, fidelity, coherence time, available gates, and topology according to the hardware \cite{Saraiva2021}. Information about these characteristics is usually provided by the service provider (although the frequency with which it is available in an updated form may vary from provider to provider).

However, what is expected to achieve quantum advantage is the ability to generate states with the necessary physical resources tied to a set of universal native gates. A given device's capabilities to fulfill this objective can vary over time. The sensitivity of qubits to noise can hinder the execution and reliability of the circuit/algorithm results. In particular, the states achieved by using the universal gate set of each device must be able to generate ``magic states'' or ``non-stabilizerness'', allowing quantum algorithms to surpass classical capabilities. 

In a 2025 paper \cite{macedo2025}, authors use Bell inequalities \cite{brunnerBellNonlocality2014} specially designed to act as witnesses for these magic states. In other previous research papers \cite{holik2024, arango2024}, quantum correlations are analyzed through different measures of entanglement by comparing universal gate sets from states generated by quantum random circuits (QRC). On the other hand, it is essential to develop testing tools that allow resource characterization without requiring developers to expose their specific algorithms, thereby preserving intellectual property or research sensitivity. The tools presented in \cite{holik2024, arango2024} are an essential step towards achieving that goal, as the characterization of correlations using quantum random circuits allows for a global evaluation of a given set of qubits.

Listing \ref{listing:qonscious-example} shows how such introspection, decision, and execution workflows look in code. The example uses Qonscious\footnote{Qonscious Framework. https://go.lifia.ar/qonscious. Last accessed in June 2025.}, a runtime framework designed to support conditional execution of quantum circuits based on resource introspection. It helps developers build quantum applications that are aware of backend conditions, such as entanglement, coherence, or fidelity, before execution. In line 23, there is a call to the \textit{run\_conditionally} function. Its first argument is an adapter on the backend where circuits are to be run. Instead of using an IBMBackend or an Aer Simulator directly, the framework defines a common interface in a family of adapters, so they are polymorphic. Moreover, when adapted backends run experiments (circuits) they all return bit-string counts, and additional experiment metadata such as timestamps, and number of shots. The second argument is a constraint on the resources that must be met by the backend. The example uses a PackedCHSHTest on a minimum value of 2.2. PackedCHSHTest runs a circuit to assess the backend ability to generate entanglement. It estimates quantum entanglement quality by running four CHSH-type measurements in parallel across four qubit pairs. It prepares Bell states, applies rotated measurements to emulate different CHSH settings, and computes correlators
$E_00$ ,$E_01$ ,$E_10$ ,$E_11$ from bitstring counts. These are combined into a CHSH score $S = E_00 + E_01 + E_10 - E_11$, which indicates the degree of non-classical correlations. Constraints are an extension point of the framework. Other constraints can be implemented and they can be combined. The third and fourth arguments are callbacks that will be called when the constraint passes or fails respectively.  These callbacks receive the backend adapter and the results of the contraint check (e.g., the result of introspection). In the example, the \textit{on\_pass} callback asks the backend to run a circuit that produces the $\Phi_+$ state and measures it. The \textit{on\_fail} callback simply reports the value of the CHSH score. 

\begin{listing}[!ht]
\centering
\begin{lstlisting}
from qiskit_ibm_runtime import QiskitRuntimeService
from qonscious.constraints import PackedCHSHTest
from qonscious.policies import MinimumAcceptableValue
from qonscious.core import IBMSamplerAdapter
from qonscious.core import executor
from qonscious.circuits import phi_plus

service = QiskitRuntimeService(channel="ibm_quantum", token='***')
backend = service.least_busy(operational=True, simulator=False)
backend_adapter = IBMSamplerAdapter(backend)

def on_pass(backend_adapter, introspection_result):
    return backend_adapter.run(phi_plus(), shots=main_circuit_shots)

def on_fail(backend_adapter, introspection_result):
    print(f"Skipping main circuit - entanglement score was {introspection_result['CHSH_score']}")

constraint = PackedCHSHTest(policy=MinimumAcceptableValue(2.2))

result = executor.run_conditionally(
        backend_adapter=backend_adapter,
        constraint=constraint,
        on_pass=on_pass,
        on_fail=on_fail,
        shots=1024
    )
\end{lstlisting}

\caption{Simple constraint checking workflow: the Bell Phi+ circuit is only executed if CHSHtest produces a value over 2.2.}
\label{listing:qonscious-example}
\end{listing}

\section{Towards Practical Runtime Resource Management}
\label{section:practical-resource-management}

While the need for a dynamic runtime layer to manage and introspect quantum hardware resources is clear, its implementation faces significant obstacles. Current quantum computing platforms impose several limitations that prevent real-time access to critical device information, constrain programmatic control over resource selection, and hinder adaptability during execution. In this section, we analyze three key challenges that must be addressed to make practical runtime resource management feasible in the NISQ era.

\textbf{Challenge 1: Limited Introspection Capabilities in Current APIs}

Most quantum programming platforms offer static access to hardware characteristics, such as qubit error rates, T1/T2 relaxation times, or gate fidelities. These parameters are typically exposed via precompiled backend snapshots (e.g., \textit{through backend.properties()} in Qiskit), and are updated only periodically by the provider. Critically, this information is not dynamically refreshed at runtime, nor can it be queried programmatically during execution.

This limitation presents a serious barrier to implementing a runtime-aware resource management layer. Without fresh introspection data, a system cannot decide whether to proceed with execution, reroute to alternative qubits, apply different circuit decompositions, or abort due to suboptimal conditions. In practice, developers must rely on outdated or average-case assumptions about device behavior. This is a significant problem in the NISQ era, where hardware variability is high and resource quality fluctuates over time.

Furthermore, most platforms lack support for programmatically specifying resource requirements as constraints or validating their availability just before execution. This requires developers to inspect static metrics and infer whether execution is feasible manually.

\textbf{Challenge 2: Extensibility and High-Level Resource Introspection}

A fundamental challenge in implementing a runtime resource management layer is the need for extensibility; that is, the capacity to incorporate new definitions of quantum resources and new introspection mechanisms as they emerge. Given the field's rapid evolution, it is unlikely that a fixed set of resource types (e.g., number of qubits, gate fidelity, T1/T2 times) will remain sufficient for long. New proposals regularly seek to quantify quantum devices' more abstract or global properties, such as non-locality, contextuality, or quantum magic.

For instance, recent work by Granda Arango et al. \cite{arango2024} demonstrates a method for analyzing how non-local correlations and multipartite entanglement are distributed across quantum states generated by noisy devices. Their approach uses quantum random circuits and statistical violation of inequalities such as Mermin and Svetlichny to characterize device capabilities in producing complex correlations. While this type of analysis is computationally expensive and requires many circuit executions and a priori knowledge of observable families, it illustrates a key idea: the capacity to introspect global quantum resources not natively exposed by the platform APIs.

This points to a broader requirement: developers should be able to define new resource metrics, plug in routines for evaluating them (either via simulation or hardware execution), and use them within a unified runtime management framework. These definitions may involve high-level properties that are not reducible to current hardware descriptors but are still relevant for deciding where, when, and how to execute a quantum algorithm.

\textbf{Challenge 3: Temporal Decoupling Between Resource Introspection and Execution}

Even when introspection routines are implemented (for example, by executing ad-hoc circuits to evaluate non-locality, entanglement, or noise levels) a critical challenge remains: the temporal decoupling between the measurement of resources and the actual use of that information.

Current quantum platforms (e.g., IBM Qiskit Runtime, Braket) operate under job-based models where circuits are queued and executed asynchronously. This means that even if a resource-checking circuit and a task-dependent circuit are submitted together (e.g., in the same Qiskit script), they are handled as independent jobs, each subject to queue delays and scheduling uncertainty.

As a result, a considerable amount of time may elapse between the introspection phase and the execution of the circuit that depends on its result. Given the fluctuations typical of NISQ devices (e.g., in coherence times, gate fidelities, or environmental noise), the resource conditions may have changed when the actual computation occurs, rendering the introspection data obsolete or misleading.

Moreover, quantum circuits do not currently support conditional branching or execution aborts at the shot level. It is impossible to write a program that aborts remaining shots mid-execution based on online measurements or partial results (which would be desirable, as pricing, for example, depends on it). This restriction severely limits the possibility of implementing reactive or adaptive logic within a single job/circuit.

Consequently, any runtime layer that aims to use high-level resource information must contend with the asynchrony and rigidity of current job execution models. Addressing this challenge will require more profound changes in quantum programming paradigms, including incorporating conditional execution models, tighter integration between classical and quantum control layers, and low-latency access to intermediate measurements.

\section{Conclusions and Future Work}
\label{section:conclusions}

In this paper, we argued that current quantum programming platforms are insufficiently equipped to handle the dynamic and variable nature of NISQ hardware. To address this gap, we introduced the concept of a dynamic runtime layer for resource management and introspection, and proposed a design that enables resource-aware quantum software development.

We identified three key challenges to practical implementation:

\begin{itemize}
    \item The limited introspection capabilities of existing APIs;
    \item The need for extensibility to accommodate evolving definitions of quantum resources;
    \item The temporal decoupling between resource measurement and dependent execution, due to rigid job-based models.
\end{itemize}

To explore how such a layer might be realized, we presented Qonscious, a proof-of-concept framework that enables conditional execution based on runtime resource checks. Qonscious is designed to be extensible, platform-agnostic, and capable of supporting high-level resource constraints, such as entanglement thresholds.

This approach opens several avenues for future work:

\begin{itemize}
    \item Extending the set of resource constraints in Qonscious to include metrics such as contextuality or circuit depth viability under decoherence limits;
    \item Supporting tighter integration with classical control loops, particularly for hybrid quantum-classical workflows;
    \item Investigating approximation or caching techniques to reduce the cost of high-level introspection routines;
    \item Exploring community standards for resource schema definition across vendors.
    \item Integrating resource-aware execution with broader service-oriented development and deployment frameworks, such as those proposed for circuit scheduling optimization \cite{alvarado-valiente_circuit_2024} and quantum service quality engineering \cite{diaz_service_2025}.
\end{itemize}

Beyond the technical aspects, we also highlight the need to foster awareness among quantum software developers regarding the nature and limitations of quantum resources. In classical software engineering, resource constraints such as memory or CPU usage are often second nature to developers. In the quantum domain, however, these constraints are less intuitive and often hidden behind layers of abstraction. Promoting a mindset that embraces resource-conscious development will be key to building reliable and sustainable quantum applications, especially in the NISQ era.

Ultimately, we advocate not only for new tools, but for a cultural shift in quantum software development, where resource awareness is embraced as a first-class engineering concern.

\bibliographystyle{IEEEtran}
\bibliography{bibliography}

\end{document}